\documentclass[a4paper,11pt]{article}
\usepackage{pos}
\usepackage{wrapfig}
\renewcommand{\logo}{\relax}

\title{Synergies of Drell-Yan, beauty, top, and Z observables in MFV-SMEFT}

\author[a]{Cornelius Grunwald}
\author[a,b]{Gudrun Hiller}
\author[a]{Kevin Kr\"oninger}
\author*[a]{Lara Nollen}

\affiliation[a]{TU Dortmund University, Department of Physics,\\ Otto-Hahn-Str.4, D-44221 Dortmund, Germany}

\affiliation[b]{Department of Physics and Astronomy, University of Sussex,\\ Brighton, BN1 9QH, U.K.}

\emailAdd{cornelius.grunwald@tu-dortmund.de}
\emailAdd{ghiller@physik.uni-dortmund.de}
\emailAdd{kevin.kroeninger@cern.ch}
\emailAdd{lara.nollen@tu-dortmund.de}

\abstract{The Standard Model Effective Field Theory (SMEFT) is a powerful tool to search for new physics in a model-independent way. 
We explore the synergies arising from different types of observables in a combined, global SMEFT fit. 
Specifically, we investigate the combination of top-quark measurements, $b\to s$ flavor changing neutral current transitions, $Z\to b \bar b$ and $Z\to c \bar c$, as well as Drell-Yan data from the LHC. 
We also examine the impact of Minimal Flavor Violation (MFV) as a flavor pattern in the global fit. 
We find that the combination of high-p$_T$ with flavor physics observables provides powerful synergies that significantly improve the fit and enable more precise tests of various SMEFT operators. 
By incorporating different observables, we are able to remove flat directions in the parameter space and infer on the flavor structure based on the MFV parameterization. 
In particular, we find that MFV significantly strengthens the constraints in comparison to a flavor-specific approach. Furthermore, our analysis yields a prediction for the dineutrino branching ratios ${\cal{B}}(B \to K^{(*)} \nu \bar \nu)$  within MFV, which can be tested experimentally at Belle II.}

\FullConference{ 21st Conference on Flavor Physics and CP Violation (FPCP 2023)\\
 29 May - 2 June 2023\\
 Lyon, France\\}

\begin{document}
\maketitle

\section{Effective Field Theories}
 The Standard Model Effective Field Theory (SMEFT) offers numerous benefits in the search for new physics (NP) above the electroweak scale. 
 Besides allowing for largely model-independent analyses, it  provides the opportunity to connect different sectors such as beauty and top via SU(2)$_L$~\cite{Bissmann:2020mfi, Aoude:2020dwv, Bruggisser:2021duo, Greljo:2022cah, Greljo:2022jac}. 
 The SMEFT Lagrangian is constructed by using the Standard Model (SM) Lagrangian as the starting point and then adding the series of all higher dimensional operators $O_i^{(d)}$ multiplied by the corresponding Wilson coefficient $C_i^{(d)}$ and suppressed by $d-4$ powers of the energy scale of BSM phsics $\Lambda$
 \begin{equation}
    \mathcal{L}_{\text{SMEFT}}\,=\, \mathcal{L}_{\text{SM}}\,+ \,\sum_{d=5}^{\infty}\,\sum_i \: \frac{{C_i^{(d)}}}{\Lambda^{d-4}}{O_i^{(d)}}  \,.
 \end{equation}
We consider only dimension-six operators, for which we employ the Warsaw basis~\cite{Grzadkowski:2010es}. 
We assume all Wilson coefficients to be real-valued as well as lepton and baryon number conserving. 
While the SMEFT is the appropriate theory for the description of collider observables, $b \to s$ observables are computed within the weak effective theory~\cite{Jenkins:2017jig}. 
Both theories are connected by RGE evolution and matching at the $W$-boson mass scale.

\section{MFV in SMEFT}

Besides reducing the number of degrees of freedom, flavor patterns impose correlations among the different flavor components of a Wilson coefficient and thus link different sectors. 
We employ the Minimal Flavor Violation (MFV) approach, in which the flavor structure of the Wilson coefficients is expanded in SM Yukawa matrix insertions, reading for example 
\begin{equation}
\label{eqn:MFV_expansion}
    \bar{q}_L q_L : \quad C_{ij} =  \left( a_1 {1}+a_2Y_uY_u^{\dagger}+a_3Y_d Y_d^{\dagger} +a_4 \left(Y_uY_u^{\dagger}\right)^2 +...  \right)_{ij} ,
\end{equation}
for an operator containing two left-handed quark doublets. 
We consider only the top quark Yukawa coupling and neglect all other Yukawa couplings in the following. 
After a rotation to the mass basis, the ansatz \eqref{eqn:MFV_expansion} yields
\begin{equation}
    \begin{split}
    C_{ij}\: \bar q_{L_i} q_{L_j} \supset \bar u_{L_i}^{\prime}\, &\left[a_1 {1}+a_2 \left[Y_u^{\text{diag}}\right]^2 +a_3 \, V \, \left[Y_d^{\text{diag}}\right]^2 V^{\dagger} +a_4 \left[Y_u^{\text{diag}}\right]^4 + ... \right]_{ij} \, u_{L_j}^{\prime}  \\ 
    + \, \bar d_{L_i}^{\prime}\, &\left[a_1 {1}+a_2\, V^{\dagger} \left[ \, Y_u^{\text{diag}}\right]^2 V +a_3 \left[Y_d^{\text{diag}}\right]^2 +a_4\, V^{\dagger} \left[ \, Y_u^{\text{diag}}\right]^4 V + ... \right]_{ij} \,d_{L_{j}}^{\prime} \,,
    \end{split}
\end{equation}
with the CKM matrix $V$. 
Powers of the top Yukwa proportional to $a_{2n}$ induce FCNCs in the down quark sector. 
Regarding the lepton sector, this approach results in lepton flavor diagonal and universal Wilson coefficients.

We define
\begin{equation}
\label{eqn:definitions}
    \tilde C_{q \bar q} = \frac{v^2}{\Lambda^2}\, \, a_1 \: , \quad \gamma_{a}=\sum_{n \geq 1} y_t^{2n}\, a_{2n}/a_1 \: .
\end{equation}
for left handed quark-doublets. 
We absorb all higher orders of top-Yukawa insertions into the ratio $\gamma_a$ as they lead to the same coupling pattern.
See~\cite{Grunwald:2023nli} for the definitions for other quark bilinears.

\section{Global SMEFT Fit}

\begin{figure}[ht]
  \centering
    \includegraphics[width=0.49\textwidth]{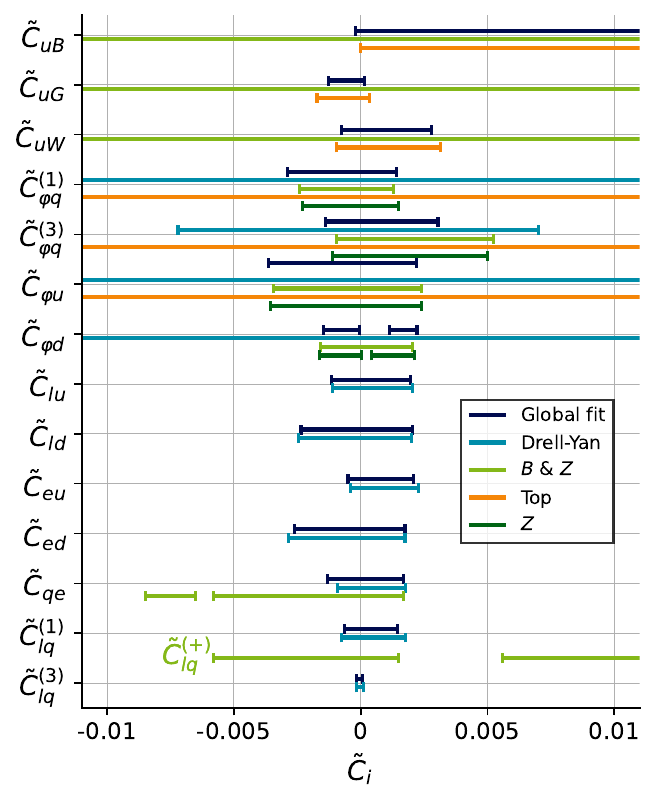}
    \includegraphics[width=0.49\textwidth]{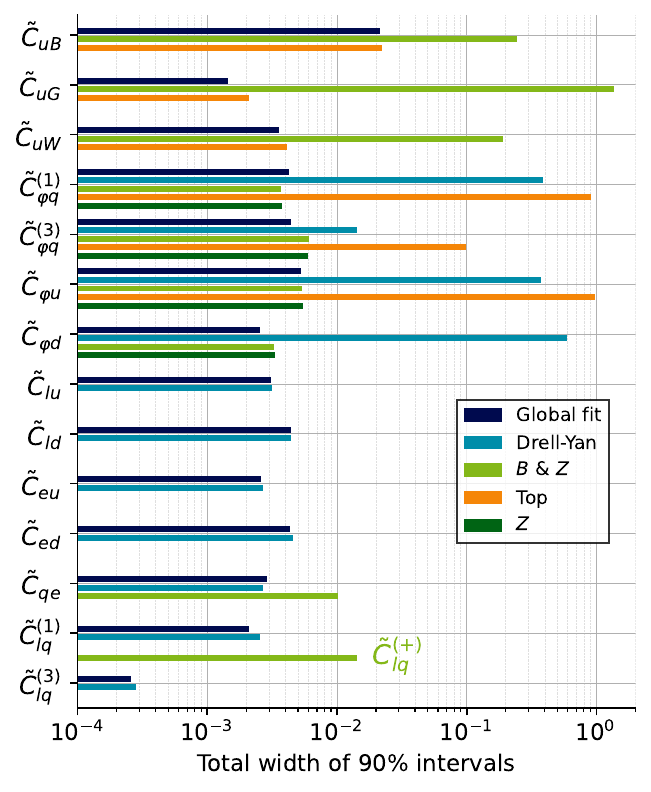}
    \caption{Constraints on the SMEFT Wilson coefficients $\tilde C_i$ assuming $\Lambda=$10\,TeV. Shown are the 90\% credible intervals (left) and the total width of these intervals (right). We compare the result of the global fit to the fit results of the individual sectors. Plots taken from Ref.~\cite{Grunwald:2023nli}.}
    \label{fig:results_wilson_coefficients}
\end{figure}

\begin{wrapfigure}{r}{0.5\textwidth}
    \centering
    \includegraphics[width=0.5\textwidth]{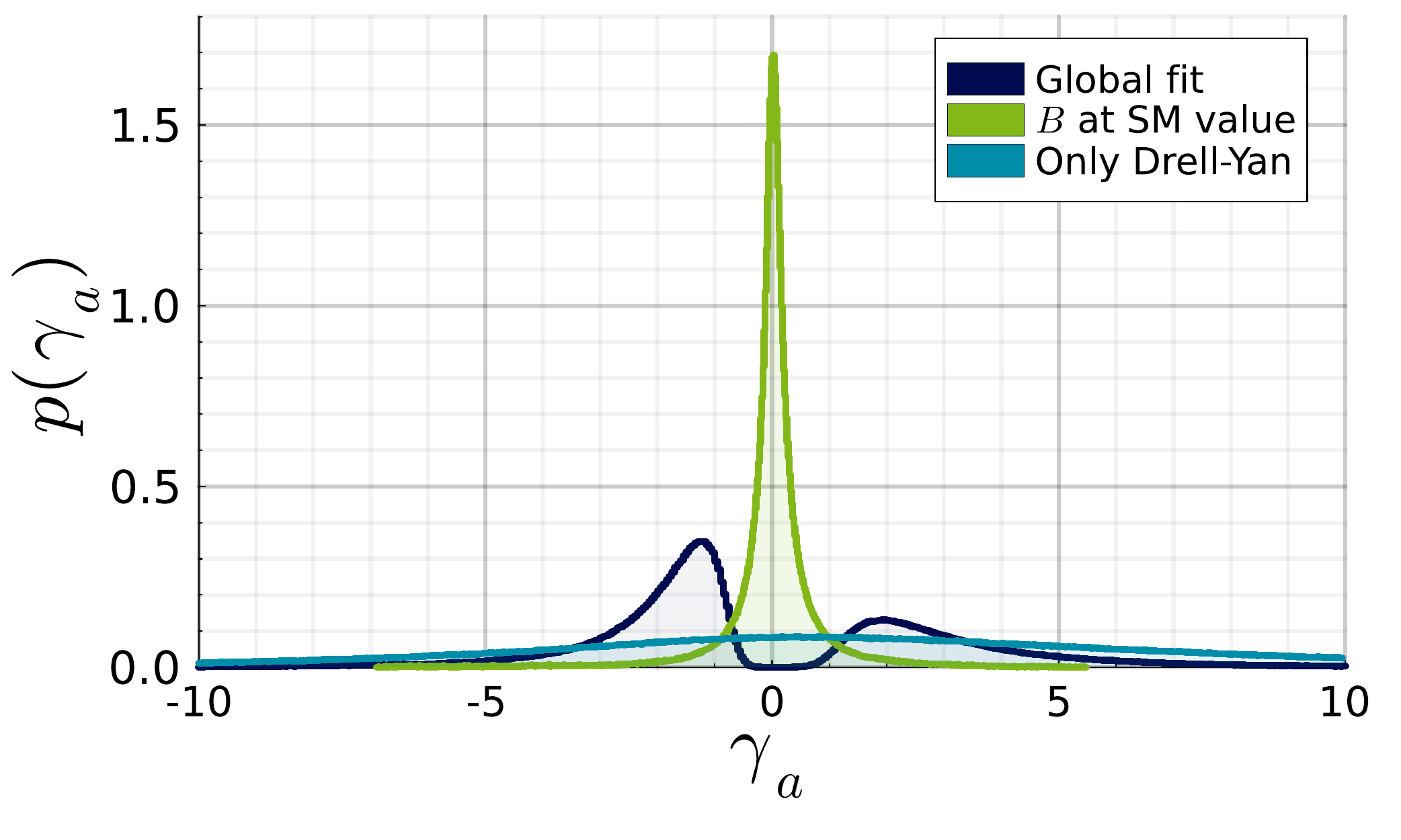}
    \caption{Marginalized posterior probability distribution of $\gamma_a$ from the global fit (dark blue), from a scenario in which all $b\to s$ measurements are set to their SM prediction (green) and from a pure Drell-Yan fit (light blue). The plot is taken from Ref.~\cite{Grunwald:2023nli}.}
    \label{fig:posterior_gamma_a}
\end{wrapfigure}

We use top-quark data, Drell-Yan measurements, $Z$ decays and $b \to s$ observables to constrain 14 Wilson coefficients and two flavor ratios $\gamma_{a/b}$, where $\gamma_{b}$ is defined for right handed up-type quarks analogously to Eq.~\eqref{eqn:definitions}. 
For the fit we employ EFT$fitter$ \cite{Castro:2016jjv}, using Bayesian inference performed with BAT.jl \cite{Schulz:2020ebm}. 
We employ a uniform prior in the range $-1\leq \tilde C_i \leq 1$ for the rescaled Wilson coefficients and a uniform prior $-10\leq \gamma_{a/b} \leq 10$ for the flavor parameters. 
Additionally, we perform fits of the individual sectors assuming a benchmark value $\gamma_{a/b}=1$. 
We choose this value as $\gamma_{a}=0$ would decouple the $b\to s$ observables. The 90\% credible intervals of the Wilson coefficients as well as the total width of these intervals are shown in Fig.~\ref{fig:results_wilson_coefficients}. 
In Fig.~\ref{fig:posterior_gamma_a}, we show the posterior probability distribution of $\gamma_a$.

We observe synergies in the global fit, which improves the bounds set by the individual sectors. 
In all fits, the constraints on the Wilson coefficients are compatible with $\tilde C_i=0$, corresponding to the SM value. 
The marginalized posterior probability distribution of $\gamma_a$ of the global fit (dark blue) shown in Fig.~\ref{fig:posterior_gamma_a} exhibits a distinct double peak behavior with a larger peak at $\gamma_a=-1.2$ and a smaller peak at $\gamma_a=1.9$, while the distribution of $\gamma_b$ (see~\cite{Grunwald:2023nli}) is rather flat. 
The fit with the $b \to s$ measurements set to their SM prediction (green), in contrast, features only one narrow peak centered around $\gamma_a=0$. 
This implies that the $b \to s$ anomalies that are currently observed at the level of $~3\sigma$ are the main driver of the shape of the posterior probability of $\gamma_a$, demonstrating that this parameter provides an interesting indirect probe of the flavor of possible physics beyond the Standard Model (BSM).

\section{Connection to dineutrino branching ratios}

Due to the manifest SU(2)$_{L}$ link in the SMEFT, $b$-hadron decays into neutrinos are directly linked to the $b$-hadron decays involving charged leptons. 
The process $b \to s \ell^+ \ell^-$ probes the combination $C_{lq}^{(+)}= C_{lq}^{(1)} + C_{lq}^{(3)} $, whereas the dineutrino process $b \to s \nu \bar \nu$ is sensitive to the orthogonal combination ${C_{lq}^{(-)}= C_{lq}^{(1)} - C_{lq}^{(3)}}$. 
A combination of both types of processes is thus crucial to pinpoint possible BSM contributions. 

\begin{wrapfigure}{l}{0.5\textwidth}
    \centering
    \includegraphics[width=0.5\textwidth]{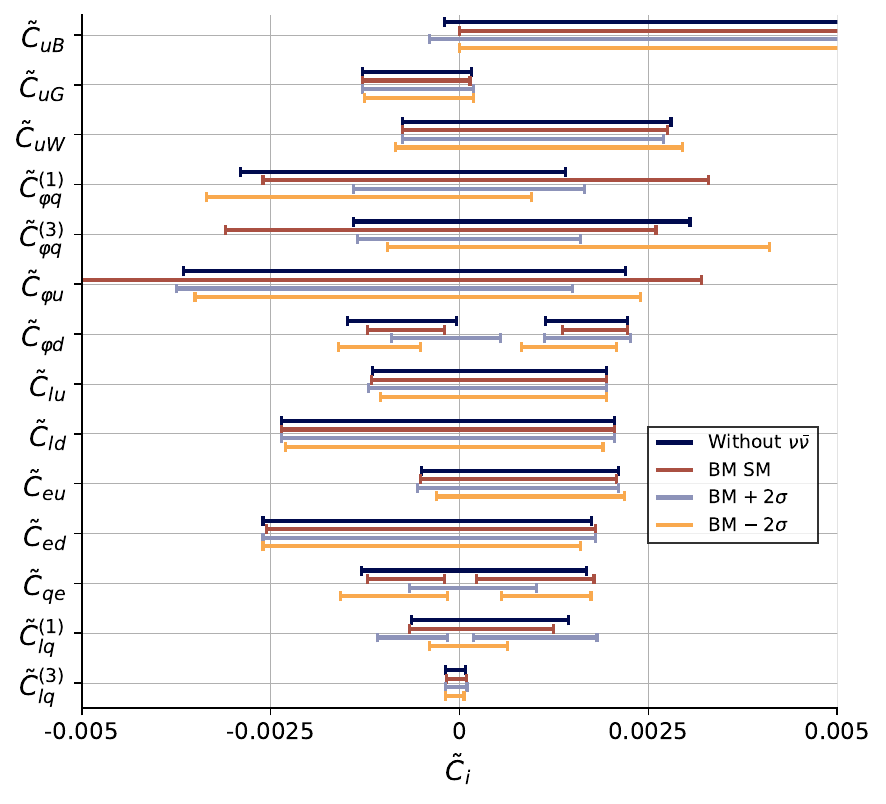}
    \caption{90\% credible intervals of the global fit with dineutrino benchmark measurements. The plot is taken from Ref.~\cite{Grunwald:2023nli}.}
    \label{fig:credible_intervals_dineutrinos}
\end{wrapfigure}

While there are currently only experimental upper limits on branching ratios of $b \to s \nu \bar \nu$ transitions, the observation of these processes is expected in the near future by Belle II~\cite{Belle-II:2018jsg}. 
To test the impact of these measurements, we perform fits including several benchmark scenarios of SM like (BM SM) as well as enhanced (BM $+2\sigma$) and suppressed (BM $-2\sigma$) branching ratios for the decays ${\cal{B}}(B^0 \to K^{*0}\nu \bar \nu)$ and ${\cal{B}}(B^+ \to K^{+}\nu \bar \nu)$. 
Both decays are sensitive to the same BSM coefficient $C_L$, as right-handed currents are suppressed in MFV by small down-type Yukawas. 
The resulting 90\% credible limits are shown in Fig.~\ref{fig:credible_intervals_dineutrinos}. 

We see that including the dineutrino measurements has a significant impact on the fit, especially on the Wilson coefficients $\tilde C_{qe}$, $\tilde C_{lq}^{(1)}$ as well as the Wilson coefficients of the penguin operators. 
Even a SM-like measurement would signal a non-zero value for $\tilde C_{qe}$ and $\tilde C_{\varphi d}$, since the inclusion of $b \to s \nu \bar \nu$ observables resolves the flat direction in the parameter space. 
Suppressed branching ratios would shift those coefficients to even larger values, whereas enhanced branching ratios would imply a non-zero value for $\tilde C_{lq}^{(1)}$. 

\begin{figure}[ht]
  \centering
    \includegraphics[width=0.49\textwidth]{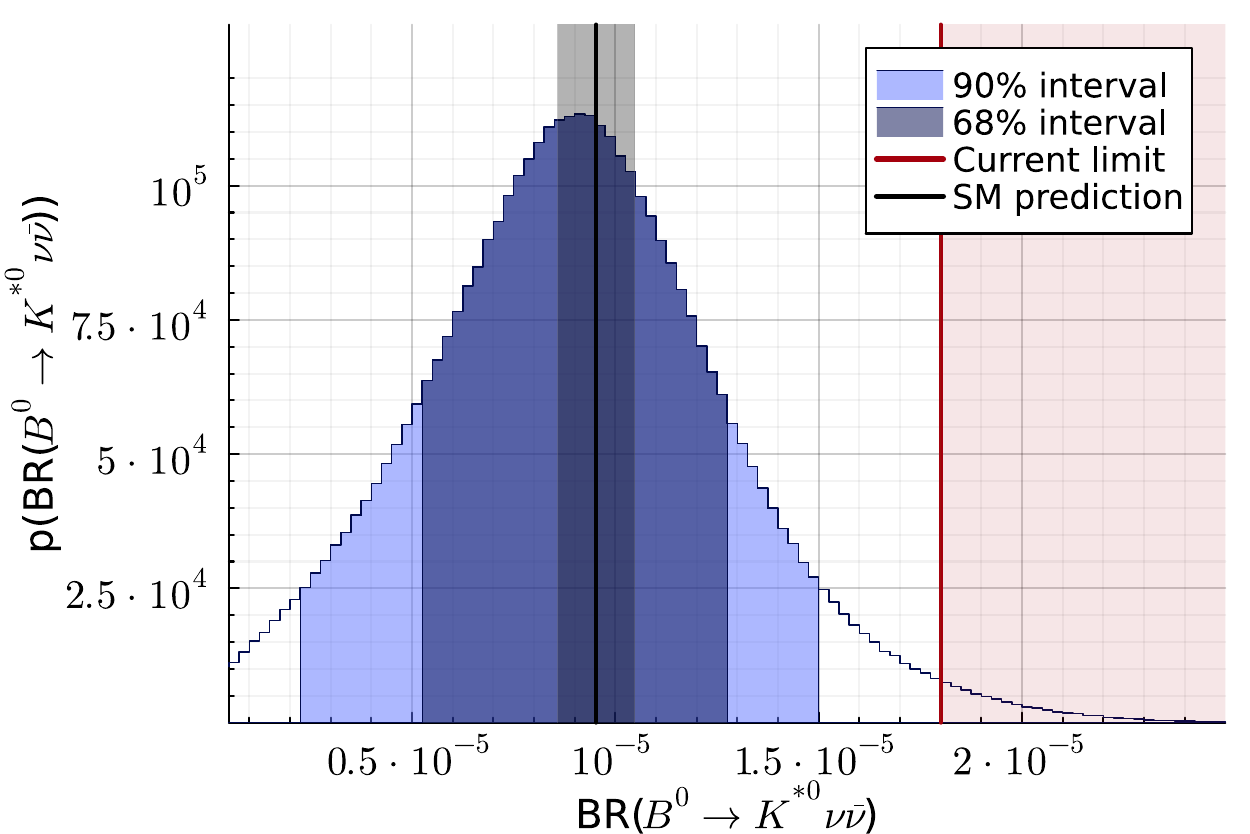}
    \includegraphics[width=0.49\textwidth]{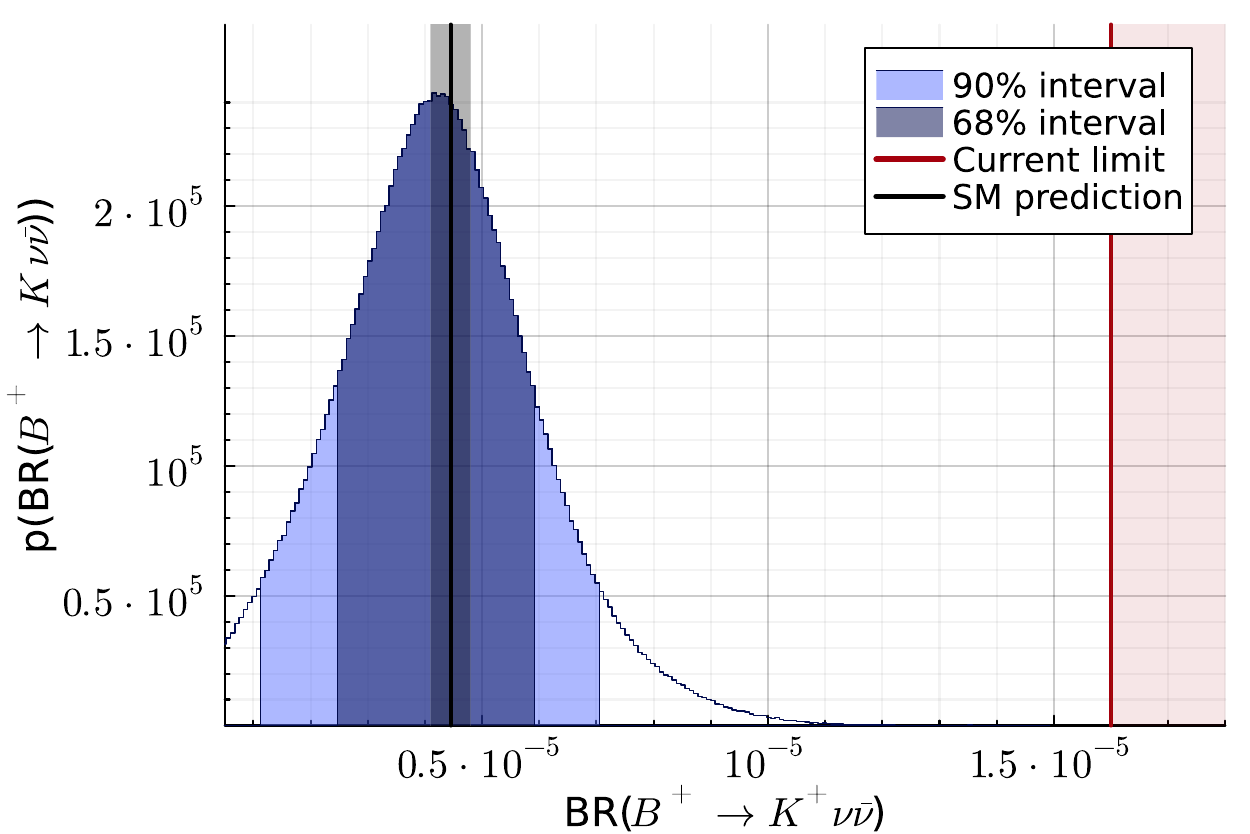}
	\caption{Marginalized posterior probability distribution of the branching ratios ${\cal{B}}(B^0 \to K^{*0}\nu \bar \nu)$ (left) and ${\cal{B}}(B^+ \to K^{+}\nu \bar \nu)$ (right). Shown are the 90\% and 68\% credible intervals in blue together with the current 90\% CL experimental limits in red as well as SM predictions with $1\sigma$ uncertainty in grey.}
    \label{fig:dineutrino_predictions}
\end{figure}

Moreover, we derive predictions for the dineutrino branching ratios based on our global fit. 
The resulting marginalised posterior probability distribution is shown in Fig.~\ref{fig:dineutrino_predictions}. 
We see that the resulting 68\% credible intervals
are centered around the SM prediction and that they are below the current experimental limit. 

\section{Conclusion}
We perform a global SMEFT fit of top, Drell-Yan, $Z$ and beauty observables within the MFV framework. 
We find that the global fit shows synergies and improves the bounds of the individual sectors. 
Presently, the global fit is consistent with the SM. 
We test the MFV expansion by including two ratios of expansion parameters as degrees of freedom in the fit. 
Our results indicate a non-zero value of $\gamma_a$, mainly resulting from the $b$ anomalies. 

We investigate the impact of future dineutrino measurements on the fit and find a significant impact on several Wilson coefficients and the potential to result in non-zero values. 
The result of the global fit is used to predict the branching ratios ${\cal{B}}(B^0 \to K^{*0}\nu \bar \nu)$ and ${\cal{B}}(B^+ \to K^{+}\nu \bar \nu)$ within the MFV scenario. 
The resulting posterior probability distributions are centered around the SM predictions, and leave sizeable room for new physics.

\section{Acknowledgments}
LN is very grateful to the organizers to be given the opportunity to present this work. LN is supported by the doctoral scholarship program of the {\it Studienstiftung des Deutschen Volkes}.

\end{document}